\documentclass[10pt,conference,letterpaper]{IEEEtran}
\usepackage{color}

\usepackage{epsfig}
\usepackage{graphicx}
\usepackage{cite}
\usepackage{amsmath}
\usepackage{amssymb}

\newcommand{\comment}[1]  {}
\def\BE{\begin{equation}}
\def\EE{\end{equation}}
\def\BEA{\begin{eqnarray}}
\def\EEA{\end{eqnarray}}

\DeclareMathOperator*{\argmax}{arg\,max}
\newtheorem{thm}{Theorem}

\newcommand{\diag}{\mathop{\bf diag}}
\newcommand{\vol}{\mathop{\bf vol}}

\newcommand{\G}{\mathcal{G}}
\newcommand{\R}{\mathbb{R}}
\newcommand{\symm}{{\mbox{\bf S}}}

\newcommand\etal{{\textsl{et al.\,\,}}}

\newcommand\va{{\bf a}} 

\newcommand\vd{{\bf d}}
\newcommand\ve{{\bf e}}

\newcommand\vg{{\bf g}}
\newcommand\vh{{\bf h}}

\newcommand\vp{{\bf p}}

\newcommand\vr{{\bf r}}

\newcommand\vw{{\bf w}}
\newcommand\vx{{\bf x}}
\newcommand\vy{{\bf y}}
\newcommand\vz{{\bf z}}

\newcommand\N{\mathcal{N}}

\begin{document}

\title{Distributed Sensor Selection \\ using a Truncated Newton Method}

\author{
\authorblockN{Danny Bickson}
\authorblockA{Machine Learning Department\\
 Carnegie Mellon University\\
Pittsburgh, PA 15213 USA\\
Email: bickson@cs.cmu.edu}
\and
\authorblockN{Danny Dolev}
\authorblockA{School of Computer Science and Engineering\\
Hebrew University of Jerusalem\\
Jerusalem 91904, Israel\\
Email: dolev@cs.huji.ac.il}
}

\maketitle

\begin{abstract}
We propose a new distributed algorithm for computing a truncated Newton method, where the main diagonal of the Hessian is computed using belief propagation. As a case study for this approach, we examine the sensor selection problem,  a Boolean convex optimization problem. We form two  distributed algorithms. The first algorithm is a distributed version of the interior point method  by Joshi and Boyd, and the second  algorithm is an order of magnitude faster approximation. As an example application we discuss distributed anomaly detection in networks. We demonstrate the applicability of our solution using both synthetic data and real traffic logs collected from the Abilene Internet backbone.
\end{abstract}

\section{Introduction}
Interior point methods are efficient algorithms for solving convex optimization problem. Typically they converge faster than gradient based methods, since they exploit information from the second derivative (the Hessian) about 
changes in the gradient to speed up convergence. However, in practice it they are not frequently deployed to large system because they involve inverting
the Hessian which has a cubic cost in the number of variables. Truncated Newton methods are an approximation where only partial information from the Hessian is used. For a good overview of interior point methods see \cite{BV04}.

In this work, we propose an efficient way for approximating the main diagonal of the Hessian inverse using
belief propagation. This provides us with a fast approximation to the interior point method, without fully inverting the Hessian matrix.  As a case study for our approach, we analyze
the sensor selection problem, which is a boolean convex optimization problem; given $m$ sensor measurements we aim at finding $k<m$ measurements that minimize the log volume of the resulting confidence ellipsoid, a scalar quantity of the uncertainty in the data. This problem is known to be NP-hard \cite{SSNP}. Recent work by Joshi and Boyd \cite{SS} proposes a solution of the relaxed sensor selection problem using an efficient interior point method. This work further gives a good overview of previous work and related algorithms in different domains.
Other work by Karuse \etal \cite{GP} discuss the related sensor placement problem,  and tackle the problem differently by minimizing the mutual information between the sensor placements selected.

Most of the existing algorithms for solving the sensor selection problem are not distributed. One of the few exception is the work of  \cite{DistSS}, which proposes a heuristic for sensor selection in sensor network context. We believe it is important to address the problem of distributed sensor selection. The  amount of data collected from sensors is rapidly growing
and centralized algorithms will not be able to process this vast amount of data. Recent survey paper by Hellerstein \cite{DBClean}, discusses the related quantitative database cleaning problem in large scale databases. As we show in the current paper, the sensor selection 
problem is closely related to the minimum volume enclosing ellipsoid problem, a technique that is commonly used to minimize the uncertainty of  stored data in large databases. Additional reason to have a distributed sensor selection algorithm is when the data is generated by a distributed set of sensors and it is not desirable 
to ship the collected data to one central processing node. In this case a distributed algorithm for selecting the ``best'' set of measurements is preferable. As a specific example of such a scenario, we discuss the problem of distributed anomaly detection in communication networks.

In the current work we extend  the previous construction of \cite{SS} by forming two distributed algorithms. The first algorithm is a distributed version of the centralized interior point method by Joshi and Boyd. The second  algorithm is a fast and light-weight distributed approximation, which is  a truncated Newton method.   An advantage of the interior point method over the heuristic algorithms is that performance bounds can be computed along the run. We demonstrate the applicability of our solution using both synthetic data and real traffic logs collected from the Abilene Internet backbone.

The structure of this paper is as follows. In Section I\ we introduce the sensor selection problem and its linear relaxation. In Section II\ we describe the Gaussian belief propagation\ algorithm, an efficient distributed iterative algorithm, which is the basis to our construction. In Section III\ we present our novel distributed algorithm, which is a distributed version of the Newton method proposed in \cite{SS}. Section IV presents our light-weight approximation for the truncated Newton method. Section V brings experimental results that compare the accuracy of both our algorithms and the original algorithm proposed in \cite{SS}. We conclude in Section VI.
 
\subsection{Problem settings}
   Suppose we are to estimate a vector from linear
measurements, corrupted by additive noise
\BE \vy = A\vx + \vw\,, \EE
where $\vx \in \R^n$ is a vector of parameters to estimate, and
$\vw \in \R^m$ is an independent identically distributed
AWGN noise $\N(0,\sigma^2I)$. We assume that $A_{m \times n}$ has full column rank $(m \ge n)$. 
The maximum likelihood estimator of $\vx$ is 
\[ \hat{\vx} = (A^TA)^{-1}A^Ty\,.\]
The $\eta$-confidence ellipsoid for $\vx-\hat{\vx}$
is the minimum volume ellipsoid that contains $\vx-\hat{\vx}$ with probability $\eta.$ The $\eta$-confidence ellipsoid is given by
\[ E_\alpha = \{\vz|\vz^T\sigma^2(A^TA)^{-1}\vz \le \alpha\}\,, \]
where $\alpha = F^{-1}_{\chi_2^n}$ is the cumulative distribution function of a $\chi^2$ random variable with $n$ degrees of freedom.
A scalar measure of the quality of estimation is the volume of the $\eta$-confidence ellipsoid:
\[ \vol(E_\alpha) = \xi\det(\sigma^{-2}A^{T}A)\,,  \]
where $\xi$ is the volume of the unit ball in $\R^n$ \cite{MVEE}. The log volume of the confidence ellipsoid gives a quantitative measure of how informative the collection of $m$ measurements is:
\[ \log \vol(E_\alpha) = \beta - \tfrac{1}{2}\log \det(A^TA)\,, \]
where $\beta$ is a constant that depends only on $\sigma,n$, and $\eta$.

Given $m$ measurements, we would like to choose a subset $n \le k < m$ of them that minimizes the log volume of the resulting confidence ellipsoid. We define the following optimization problem:
\BE \max_{\vz \in\{0,1\}^n} \log \det(A^T\diag(\vz)A)\,, \label{SS-primal} \EE
\vspace{-6mm}
\[ \text{such that} \ \ \ \ \vz^T\mathbf{1}  = k  \,,\ \ \ \ k < m\,. \]
where $\mathbf{1}$ is the appropriately sized all ones vector. This problem is known to be a boolean convex problem \cite{SS}.

\begin{table}[htb!]
\begin{center}
\footnotesize
\begin{tabular}{lcl}
  \hline
  Given & &  feasible starting point $\vx_0$ and tolerance $\epsilon > 0$, $k = 1$  \\
  \hline
  Repeat & 1 & Compute the Newton step and decrement \\
  & & $\Delta \vz = -H^{-1}\vg + (\frac{\mathbf{1}^TH^{-1}\vg}{\mathbf{1}^TH^{-1}\mathbf{1}})H^{-1}\mathbf{1}$, \ \ \ \ \ $\lambda^2 = g^{T}  \Delta \vz;$ \\
   & 2 & Stopping criterion: quit if $ \lambda^2/2 \le \epsilon;$\\
   & 3 & Line search: Choose step size t by
backtracking line search; \\
  & 4 & Update: $\vx_k := \vx_{k-1} + t\Delta \vz, \ \ \ k = k + 1.$ \\
  \hline
\label{tab:Newton}
\end{tabular}
\end{center}
\caption{The constrained Newton algorithm \cite[$\S 10.2.2$]{BV04}
.}
\label{newton}
\end{table}

\subsection{Relaxed problem}
Recent work by Joshi and Boyd \cite{SS} proposes to relax \eqref{SS-primal} to allow a fractional $0 \le z_i \le 1$ to get a relaxed version of the problem:
\BE \max_{\vz \in [0,1]^n} \log \det(A^T\diag(\vz)A)\,, \label{eq:relaxed}\EE
\[ \text{such that} \ \ \ \ \vz^T\mathbf{1}  = k\,,\ \  k < m\,. \]
A common technique for solving the relaxed problem is the log barrier method. In the log barrier method the constrains $0 \le z_i \le 1$ are incorporated into the cost function
\BE \max_{\vz \in [0,1]^n} \log \det(A^T\diag(\vz)A) +\kappa\sum_{i=1}^m(\log(z_i)+\log(1-z_i))\,, \label{eq:SS-approx}\EE
\[ \text{such that} \ \ \ \ \vz^T\mathbf{1}  = k\,, \ \ \ k < m\,, \]
where $\kappa > 0$ is a weighting parameter that controls the accuracy of the approximation. The approximated function \eqref{eq:SS-approx} is concave
and smooth and thus it is possible to use the Newton method \cite[\S10.2.2]{BV04} efficiently.
Table \ref{tab:Newton}
outlines the constrained Newton method. Starting from an initial feasible point $\vx_0$, at each Newton step we compute the search direction $\Delta \vz$. Then a backtracking line search \cite[\S9.2]{BV04} is used to compute a step size $t$. The current solution $\vx_{k}$ is replaced by  $\vx_{k} + t\Delta \vz$. We stop when the Newton decrement is small. 

For computing $\Delta \vz$ we need to compute the Hessian $H$ and the gradient $\vg$. The gradient is given by
\BE \vg = -\diag(AXA^T)  +\kappa \vh\,, \label{eq:gradient} \EE
where \[ X = (A^T\diag(\vz)A)^{-1},\ \   h_i = 1/z_i + 1/(1-z_i),\ \   \forall i = 1 \dots m\,. \]

The Hessian is given by
\BE H=-(AXA^T)\circ(AXA^T)-\diag(\kappa\vp)\,, \label{eq:Hessian} \EE where $\circ$ is the Hadamard (elementwise) product, \[ p_i = 1/z_i^2+1/(1-z_i)^2,\ \  \forall i \ = 1 \dots m\,.\]
Finally the search direction $\Delta z$ is computed by 
\BE \Delta \vz = -H^{-1}\vg + (\frac{\mathbf{1}^TH^{-1}\vg}{\mathbf{1}^TH^{-1}\mathbf{1}})H^{-1}\mathbf{1}\,. \label{eq:deltaZ} \EE

\section{Gaussian belief propagation}
\label{sec:GaBP}
We wish to compute the \emph{maximum a posteriori} (MAP) estimate
of a random vector $x$ with Gaussian distribution (after conditioning
on measurements):
\begin{equation}
p(\vx) \propto \exp\{ -\tfrac{1}{2} \vx^T J \vx + \vh^T \vx \}\,, \label{eq:sys_prob}
\end{equation}
where $J \succ 0$ is a symmetric positive definite matrix (the information
matrix) and $\vh$ is the potential vector.  This problem is equivalent to solving
$J \vx = \vh$ for $\vx$ given $(\vh,J)$ or to solve the convex quadratic
optimization problem:
\begin{equation} \label{eq:objective_func}
\mbox{minimize} \,\, f(x) \triangleq \tfrac{1}{2} \vx^T J \vx - \vh^T \vx\,.
\end{equation}
GaBP is an efficient
distributed message-passing algorithm for inference over a
Gaussian graphical model. Given the Gaussian density function \eqref{eq:sys_prob} or objective function \eqref{eq:objective_func},
we are interested in calculating the
marginal densities, which must also be Gaussian,
\BE p(x_{i}) \sim
\mathcal{N}(\mu_{i}=(J^{-1} h)_{i},K_i \triangleq (J^{-1})_{ii})\,,
\nonumber \EE where $\mu_{i}$ and $K_{i}$ are the
marginal mean and variance,
respectively. The GaBP update rules are summarized in
Table~\ref{tab_summary}. We write $\mathbb{N}(i)$ to denote 
the set of neighbors of node $i$ (non zero entries in $J$ at row $i$. 
\footnotesize
\begin{table}
\centerline{ 
\begin{tabular}{|c|c|l|}
  \hline
  \textbf{\#} & \textbf{Stage} & \textbf{Operation}\\
  \hline
  1. & \emph{Initialize} & Set $\alpha_{ij}=0$ and $\beta_{ij}=0$, $\forall (i,j) \in \G$\\ \hline
  2. & \emph{Iterate} & For all $(i,j) \in \G$\\
  & & \, $\alpha_{i \backslash j} = J_{ii} + \sum_{{k} \in \mathbb{N}(i) \backslash j} \alpha_{ki}$\\
  & & \, $\beta_{i \backslash j} = h_i + \sum_{k \in \mathbb{N}(i) \backslash j} \beta_{ki}$\\
  & & \, $\alpha_{ij} = -J_{ij}^2 \alpha_{i \backslash j}^{-1}$ \\
  & & \, $\beta_{ij} = -J_{ij} \alpha_{i \backslash j}^{-1} \beta_{i \backslash j}$\\
  & & end \\ \hline
  3. & \emph{Check} & If $\alpha$'s and $\beta$'s have converged,\\
  & &  continue to \#4. Else, return to \#2.\\\hline
  4. & \emph{Infer} & $\hat{K}_{i}=(J_{ii} + \sum_{{k} \in \mathbb{N}(i)} \alpha_{ki})^{-1}$ \\
& &  $\hat{\mu}_{i}= \hat{K}_i (h_i + \sum_{k \in \mathbb{N}(i)} \beta_{ki})$.\\
  \hline
  5. & \emph{Output} & $x^*_{i}= \hat{\mu}_{i}, \forall i.$ \\\hline
\end{tabular}}
\vspace{0.5cm}
\caption{Computing $\vx^*  = \argmax_{\vx} \exp(-\tfrac{1}{2}\vx^TJ\vx + \vh^T\vx)$ via GaBP.} \label{tab_summary}
\vspace{-8mm}
\end{table}

\normalsize

It is known that if GaBP converges, it results in the exact
MAP estimate $x^*$, although the variance estimates $\hat{K}_i$ 
computed by GaBP are only approximations to the correct
variances \cite{BibDB:Weiss01Correctness}. 
\section{First algorithm: distributed interior point method}
Following our previous work \cite{ISIT09-3,Allerton08-1}, we propose a way to distribute the 
Newton method for solving the sensor selection problem. This construction serves mainly for 
comparing the computational overhead and the accuracy with our approximation algorithm, 
presented in the next section. There are several challenges that make the sensor selection 
problem harder to distribute than our previous work \cite{Allerton08-1} that addressed 
a distributed version of standard linear programming. First, the computation of the gradient, \eqref{eq:gradient}, 
involves inverting a matrix $Q$ and computing the full inverse, which is needed for computing 
the Hessian. Furthermore, the search direction $\Delta z$ computation, \eqref{eq:deltaZ}, 
requires solution of two systems of linear equations.
Typically in linear programming the computation of the search directions involves a solution of 
single linear system of equations and not the full inversion of the Hessian matrix. Second, the 
linear iterative algorithms, which are used for efficiently solving the linear system of equations 
empirically, failed to converge. To this end, we deploy our recent construction \cite{ISIT09-1} 
for forcing convergence of the iterative algorithms to the correct solution.

Note that the distributed algorithm is not an approximation, it solves the relaxed sensor selection problem \eqref{eq:relaxed} accurately. However, there is higher computational cost relative to the approximation algorithm presented in next section.\subsection{Distributed computation of the gradient}
At each Newton step, we would like to compute the following gradient
\small
\[ \vg = -\diag(A(A^T\diag(\vz)A)^{-1}A^T)  +\kappa \vh\,. \]
\normalsize
We propose to utilize the GaBP algorithm for computing the matrix inverse $(A^T\diag(\vz)A)^{-1}$ using the following construction. Denoting $Q=(A^T\diag(\vz)A)$, we solve $n$ instances of linear systems of equations, where the $i$-th system, $Q\vr_{i} = \ve_i$, $\vr_{i} \in \R^n$, is the solution, and $\ve_i$ is a vector with $1$ at the $i$-th position and zero elsewhere. This is equivalent to computing $Q^{-1}$, since for each $i$ we compute
$\vr_i =Q^{-1}\ve_i$, which is the $i$-th row of the required solution. Note that the computation can be done in parallel, since the $i$-th equation does not depended on the solution of the other equations. To distribute this computation, each computing node gets one row of the matrix $Q$ and the matching entry of the vector $\ve_i$. It is known that when the GaBP\ algorithm converges it converges to the correct solution, so $Q^{-1}$ is not an approximation. After computing $Q^{-1}$ the gradient is computed by multiplying by $A(Q^{-1})A^T$, a computation that is fairly easy to distribute. Finally, the diagonal entry is selected, $\diag(A(Q^{-1})A^T)$, and each computing node adds the scalar $\kappa h_i$ to its matching gradient entry. The result of this computation is that  $g_i$ is stored in the $i$-th computing node.\ \subsection{Computing the Hessian and search direction}
After computing the gradient, each computing node has the matching row of $AXA^T$.\ The Hessian is computed by multiplying the matrices $AXA^T$ using the elementwise product: $H = -(AXA^T)\circ(AXA^T)-\diag(\kappa\vp)$. The result of this operation is that each computing node has the matching row in the Hessian. 

For computing the search direction $\Delta z$, \eqref{eq:deltaZ}, two linear systems of equations are  solved: $H^{-1}\vg$ and $H^{-1}\mathbf{1}.$  Again, we use the GaBP algorithm for computing these solutions. These two computation can be done in parallel, as well. 

\subsection{Convergence and overhead}
We use the recent construction of Johnson \etal \cite{ISIT09-1} for forcing convergence of a 
linear iterative algorithm to the correct solution. The construction is applied in 
three places: in computing the inverse matrix $Q^{-1 }$and in solving $H^{-1}\vg$ and 
$H^{-1}\mathbf{1}$. 

Next we discuss the overhead of the distributed Newton method. Typically there are 
around 10-20 Newton iterations until convergence. In each Newton step, in the gradient 
computation the inversion of the matrix $Q$ requires solving $n$ linear systems of equations 
(which can be solved in parallel). In theory, a logarithmic number of iterations is required for 
convergence. An upper bound on converge speed is presented in~\cite{Allerton08-1}. In total, 
the cost of inverting $Q$ is $O(n^3\log(n)).$ Note that this cost is higher than the traditional 
Gaussian elimination $O(n^3)$, however we require only $O(\log(n))$ communication rounds, relative 
to $n$ communication rounds that are needed for a distributed implementation of Gaussian 
elimination.  The computation of the search direction involves a solution of two linear systems 
of equations with dominating overhead of $O(m^2\log(m))$. 

The above analysis is for the worst case. In case the matrix $Q$ is sparse, further speedup can be obtained. In practice, fast convergence (tens of iterations)\ of the GaBP\ algorithm was observed in problem sizes of millions of variables\cite{phd-thesis}. Note, as mentioned above,  the distributed Newton method is presented here mainly for reference.

\section{Second algorithm: fast approximation algorithm}
Our goal is to allow a distributed and efficient computation of \eqref{eq:SS-approx} for solving the approximate problem. We propose an approximate computation (truncated Newton method)\ that is composed of two steps: approximation of the gradient computation \eqref{eq:gradient}, and an approximation of the Hessian \eqref{eq:Hessian}.
In Section \ref{sec:exp_results} we show that our approximation is comparable in performance to the original centralized algorithm. 
\subsection{Approximation of the gradient}
\begin{thm}
\label{thm1}
The gradient, \eqref{eq:gradient}, can be computed by inverting the following matrix: \BE \label{eq:E} E = \left(
  \begin{array}{cc}
    \mathbf{0} & A^T \\
    A & Z \\
  \end{array}
\right)\,,\EE  and taking the diagonal of the lower right block, where $Z=-\diag(\vz)^{-1}$.

\end{thm}
\begin{proof}
Using the Schur complement of the matrix
$M = \left(
  \begin{array}{cc}
    A & B \\
    C & D \\
  \end{array}
\right)$ the lower right block of $M^{-1}$ is \small \[ Y = D^{-1} +D^{-1}C(A-BD^{-1}C)BD^{-1}. \]\normalsize Substituting $A \triangleq \mathbf{0}, B \triangleq A^T, C \triangleq A, D \triangleq -Z^{-1}$ we get \small \BE Y = -Z^{}+Z^{}A(\mathbf{0}+ A^TZA)^{-1}A^TZ\,. \label{eq:Y} \EE  \normalsize Multiply by $Z^{-2}  $ to get $Z^{-2}Y=Z^{-1}+A(A^TZA)^{-1}A^T.$ Equivalently, $A(A^TZA)^{-1}A^T=Z^{-2}(Y+Z).$
Finally the gradient is given by
\small
\BE \vg = -\diag(Z^{-2}(Y+Z))  +\kappa \vh \label{eq:gradient-approx}\,. \EE
\normalsize
\vspace{-1mm}
\end{proof}
Theorem \ref{thm1} shows an alternative way of accurately computing the gradient
in each Newton step. The following theorem presents how to compute \eqref{eq:gradient-approx} approximately.
\begin{thm}
The gradient, \eqref{eq:gradient-approx}, can be computed approximately and distributively using the GaBP algorithm. 
\end{thm} 
\begin{proof}
Given the matrix $E$ defined in \eqref{eq:E} we construct the following 
multivariate probability distribution \[ p(\vx) = \exp(-\tfrac{1}{2}\vx^TE\vx+\vd^Tx)\,. \]
It is shown in\cite{BibDB:Weiss01Correctness} that the GaBP algorithm, when converging, computes the exact MAP assignment of $\vx$, which is the vector $E^{-1}\vd$, as well as an approximation to the variance estimates $K_i \approx\ \diag(E^{-1})$.
We are interested only in the $m$ last entries of $K_{n+1} \dots K_{n+m+1} $, which is an approximation 
to the main diagonal of $Y$ as defined in \eqref{eq:Y}. We denote   this approximation (the output of the  GaBP algorithm) as $\tilde{Y} \triangleq(K_{n+1} \dots K_{n+m+1}) \in \R^m$. Now we can compute the approximated gradient:
\BE \tilde{\vg} = -\diag(Z^{-2}(\tilde{Y}+Z))  +\kappa \vh \label{eq:gradient-approx}\,. \EE
\vspace{-1mm}
\end{proof}
\vspace{-0mm}
\subsection{Approximating the Hessian and computing the search direction}
We propose to use a truncated Newton method, where only the main diagonal of the Hessian is computed.
\[ \tilde{H}= - \diag(\tilde{\vg}\circ\tilde{\vg})-\diag(\kappa\vp)\,. \]
Next, we need to compute the search direction (step 1 in Table \ref{tab:Newton}): \[ \Delta \vz = -\tilde{H}^{-1}\vg + (\frac{\mathbf{1}^T\tilde{H}^{-1}\vg}{\mathbf{1}^T\tilde{H}^{-1}\mathbf{1}})\tilde{H}^{-1}\mathbf{1}\,. \] 
Since $H$ is diagonal, this computation can be done in $O(m)$.
Note that our construction for approximating the Hessian is general and can be used in other problems as well.

\subsection{Cost analysis}
\begin{table}[h!]
\small
\centering{
\begin{tabular}{|l|c|c|}\hline
Algorithm & Sensor Selection \cite{SS} & Fast Approximation \\\hline \hline
Gradient computation & $O(n^3)$ & $O(mn\log(m+n)$) \\\hline
Hessian computation & $O(m^2)$ & $O(m)$ \\\hline
Search direction & $O(m^2)$ & O(m) \\\hline
Total cost & $O(n^{2 } +m^3)$ & $O(mn\log(m))$ \\\hline
\end{tabular}
}
\caption{Algorithm cost comparison}
\vspace{-6mm}
\end{table}
\normalsize
Computing the Newton step in \cite{SS} is dominated by inverting $A^T\diag(\vz)A$ in the 
gradient  computation, which costs $O(n^3)$. Computing the Hessian involves computing the 
product $(AXA^T)\circ(AXA^T)$, which costs $O(m^2)$. Computing the search direction $
\Delta \vz$ costs another $O(m^2)$. In total we get $O(n^{2 } +m^3)$. In contrast, in our 
method we compute an approximation of the gradient by running the GaBP algorithm. 
Assuming $A$ is dense, we get $O(mn)$ non-zero elements in the matrix $E$.
When the GaBP algorithm converges, the number of iterations is typically logarithmic in the 
matrix size (convergence rate analysis can be found on \cite{Allerton08-1}). The Hessian is 
computed by its diagonal approximation, which costs only $O(m)$. The search direction 
computation takes another $O(m)$. In total the heaviest operation is the gradient 
computation, which dominates the total complexity. It is important to note that our method is 
an approximation to the method proposed in  \cite{SS}, where its efficiency is discussed. 

\section{Relation to the minimum volume enclosing ellipsoid problem\\ }
\label{s-MVEE}
The minimum volume enclosing ellipsoid problem (MVEE) is formulated as follows. Given $m$ vectors of dimension $n$, we aim at finding the smallest ellipsoid that encloses all vectors:
\BE 
\min_{M \in \symm^n}-\log \det M\,,
\EE
\[  \mbox{such that  \ \ \ } \va_i^TM\va_i \le 1  \ \ \ \ \ \forall i=1,\dots,m\,,
 \]
 where $\symm^n$ is the set of symmetric $n \times n$ matrices. The following theorem establishes the connection between the dual of the MVEE\ and the sensor selection problem.
This means that if we have an efficient algorithm for computing the sensor selection problem we can also solve the MVEE problem efficiently. The relation to the dual was also observed in~\cite{SS}, but here we provide an alternative and simpler proof. 
\begin{thm}
The dual of the MVEE is related to the relaxed sensor selection problem, \eqref{eq:relaxed}, where the number of selected sensors is $k=n$. \\
\begin{proof}
We start by forming the Lagrangian
\BE L(M,\lambda) = -\log \det M + \sum_{i=1}^n z_i(1-\va_i^TM\va_i)\,, \label{eq:lagrangian}\EE
where $\vz\ge0$ is a vector of Lagrange multipliers.
Now we compute the optimality conditions by computing the derivative and comparing it to zero.
\small
\BEA
\frac{\partial L(M,\lambda)}{\partial M} = -M^{-1}+A^T\diag(\vz)A = 0\,,  \nonumber \\ M^{-1} = A^T \diag(\vz)A\,, \ \ \ \ \ \ \ \label{eq:M}\\
\frac{\partial L(M,\lambda)}{\partial \vz} =\sum_{i=1}^{n}(\va_i^TM\va_i-1) =0\,,\nonumber\\
\sum_{i=1}^n\va^T_iM\va_i=n\,. \ \ \ \ \ \ \ \ \ \label{eq:dz}
\EEA
\normalsize
The dual of the MVEE\ problem is obtained by substituting \eqref{eq:M} and \eqref{eq:dz} into \eqref{eq:lagrangian} to get
\small
\[ \mathop{\max \log \det} \limits_\vz A^T \diag(\vz)A\,, \]
\[ \mbox{such that \ \   } \vz\mathbf{1} = n\,, \ \ \ \vz\ge0\,. \]
\normalsize
Notice the relation to \eqref{eq:relaxed}, where the number of sensors is $k=n$.
The only difference is that in the sensor selection problem $\vz \in [0,1]^n,$ whereas here $\vz\ge0$. \end{proof}
\end{thm}

\section{Experimental results}

\label{sec:exp_results}
We evaluate our proposed algorithms using two datasets. The first one is a synthetic dataset, which is borrowed from \cite{SS}, and the second one is real data acquired from the Abilene Internet2 backbone network\cite{Abilene}. Our simulation is heavily based on the Matlab simulation of \cite{SS}, and is available on the web \cite{MatlabGABP}.
In all experiments we have used a Newton tolerance of $10^{-3}$ and the GaBP threshold was set to $10^{-8}$.
\subsection{Synthetic data}Following \cite{SS} we use $m=100$ potential sensors and $n=20$ parameters to estimate. The $m$ measurement vectors $\va_i \in \R^n$ are chosen randomly,
and independently, from the  distribution $\N(0,\frac{1}{\sqrt{n}}I)$.
We solve the relaxed problem, \eqref{eq:relaxed}, with distributed Newton construction as well as our distributed approximate algorithm (denoted as GaBP in the graphs). Figure 1 shows the duality gap of the different methods tested. The $y$-axis presents the duality gap of the cost function, the $x$-axis ticks are the number of sensors selected. As predicated, the distributed Newton method has performs better than GaBP, because GaBP computes an approximation to the relaxed problem. However, when deploying the local search heuristic proposed in \cite{SS}
we get a very good solution, which outperforms the Newton method. 

Figure 1 (top)\ plots some bounds on the solution quality, where upper bound is computed using the fact that the relaxed sensor selection problem solution is at most $2m\tau$ from the optimal solution \cite[\S10.4]{BV04}, the lower bound is computed
using the simple selection rule. Improved lower bounds
are obtained via the local optimization procedure. A simple method to
carry this local optimization is to start from the $k$ selected sensors, and check sensor selections that can be derived from by swapping one of the chosen sensors with one of the $m-k$ sensors not chosen.

Typically the number of Newton iterations in both algorithms are $4$-$8$. As expected, the approximated algorithm converged very rapidly, taking 9-10 iterations for each Newton step. In contrary, the full Newton method required 20-50 GaBP iterations for each row of $Q^{-1}$ (those iteration can be done in parallel by increasing message sizes in the network). The convergence enforcement construction used in the Newton method required up to 25 iterations of the outer loop. To conclude, the approximated algorithm requires tens of iterations, while the accurate distributed Newton method requires hundreds of iterations.

\begin{tabular}{c c}
\hspace{-0cm}
\includegraphics[scale=0.2,clip,bb=22 179 611 617]{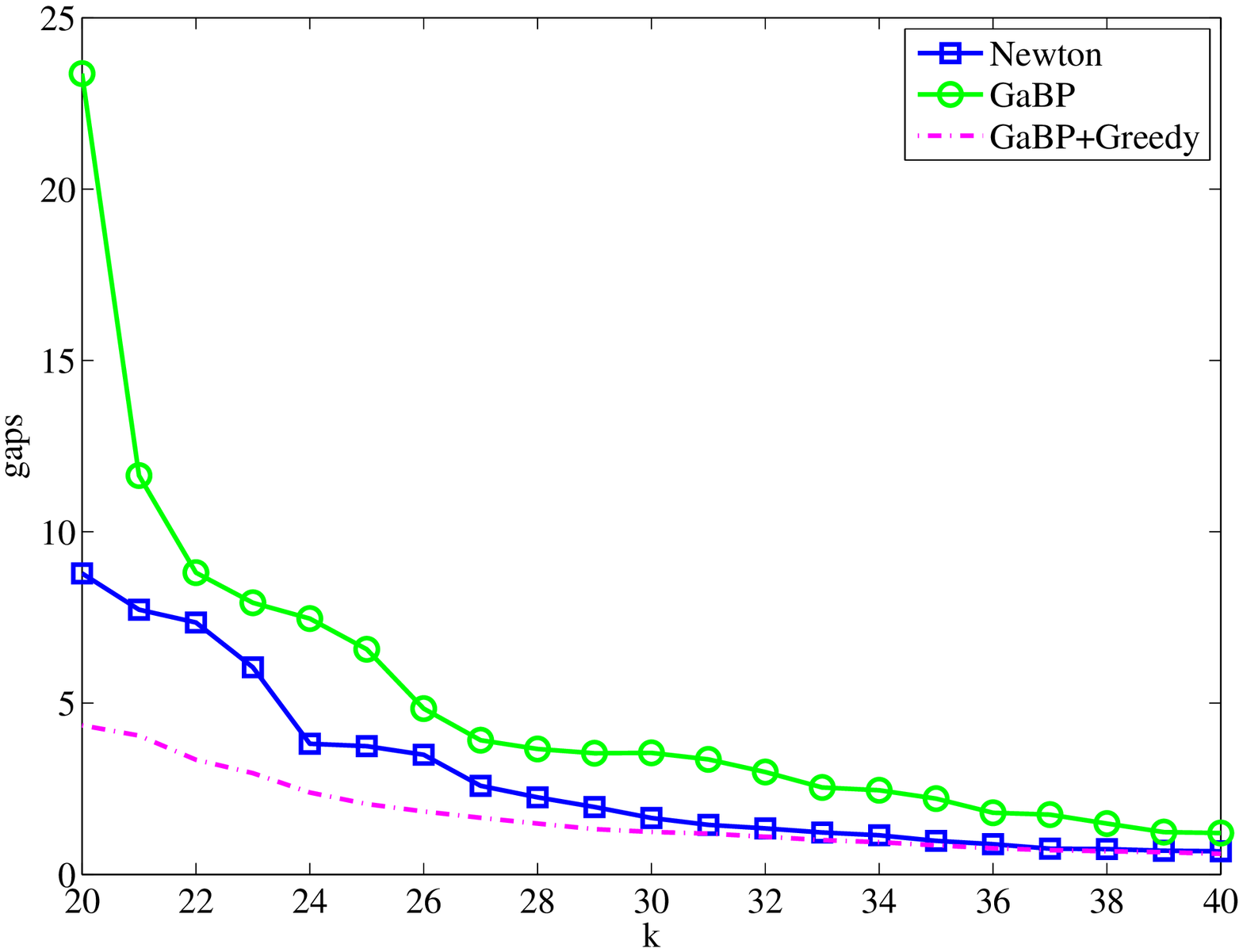}
&
\hspace{-0.6cm}
\includegraphics[scale=0.2,clip,bb=13 175 613 615]{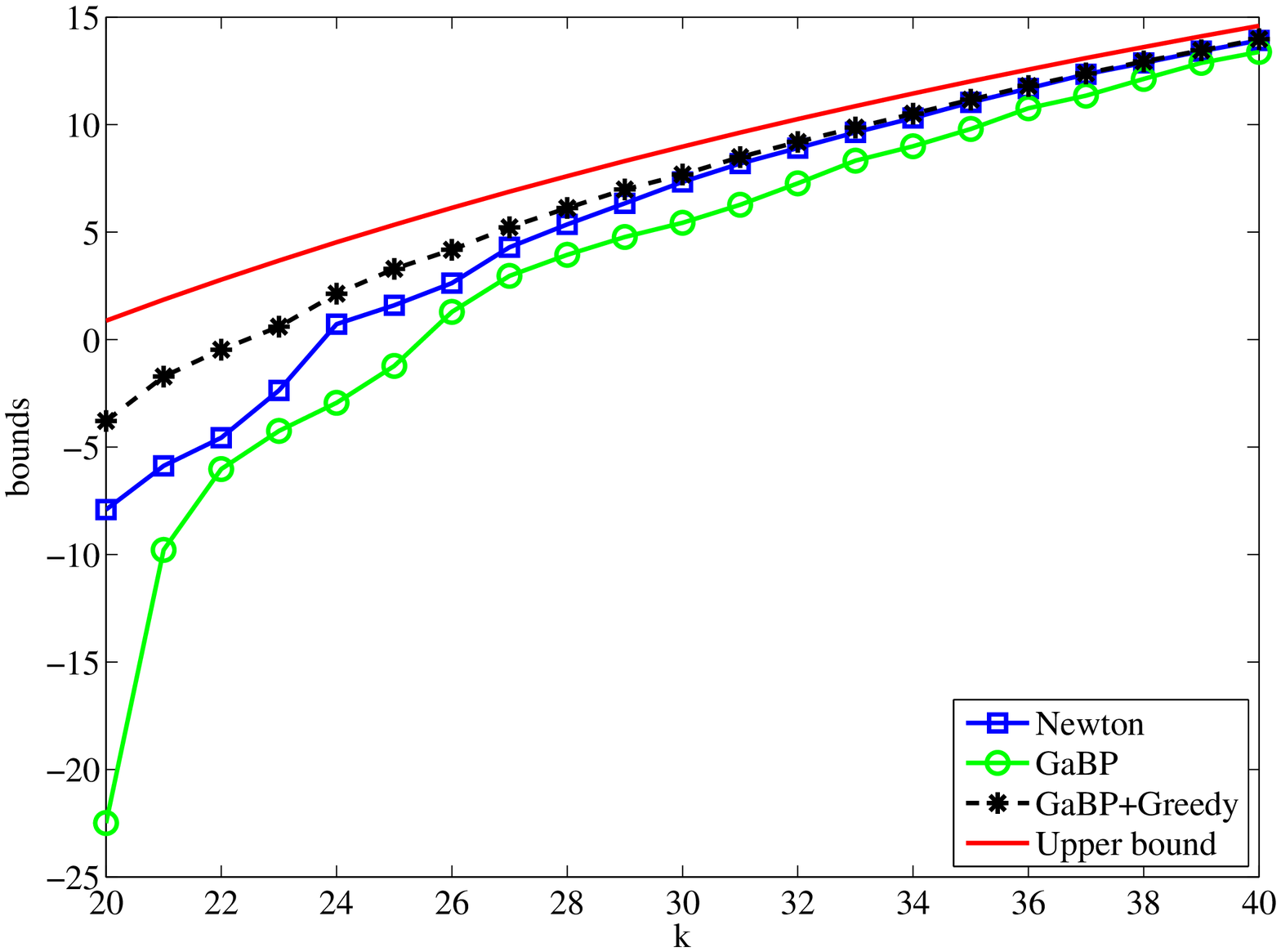}\\
\hspace{-0cm}
\includegraphics[scale=0.2,bb=41 168 593 605]{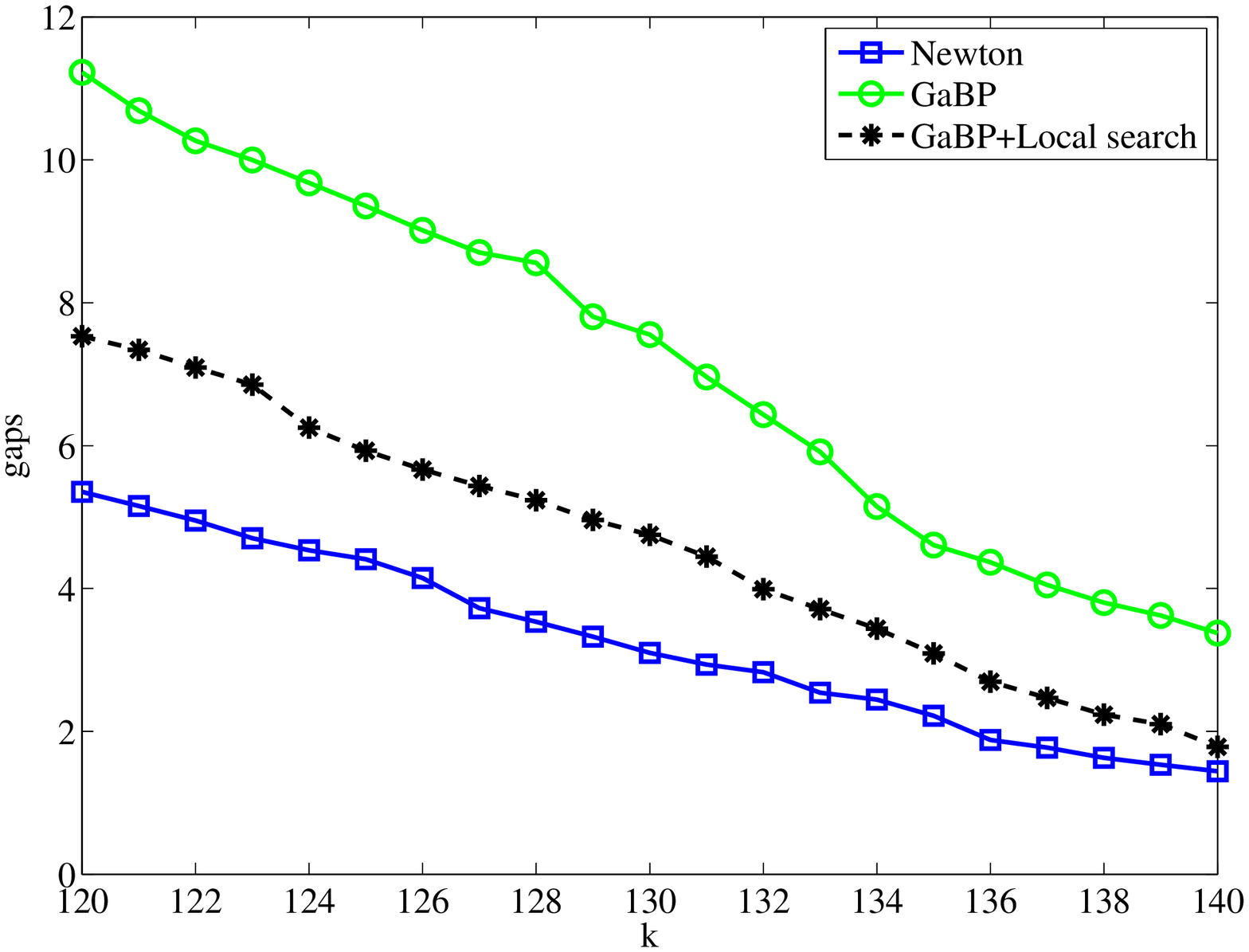}
&
\hspace{-0.6cm}
\includegraphics[scale=0.2,bb=22 179 611 617,clip]{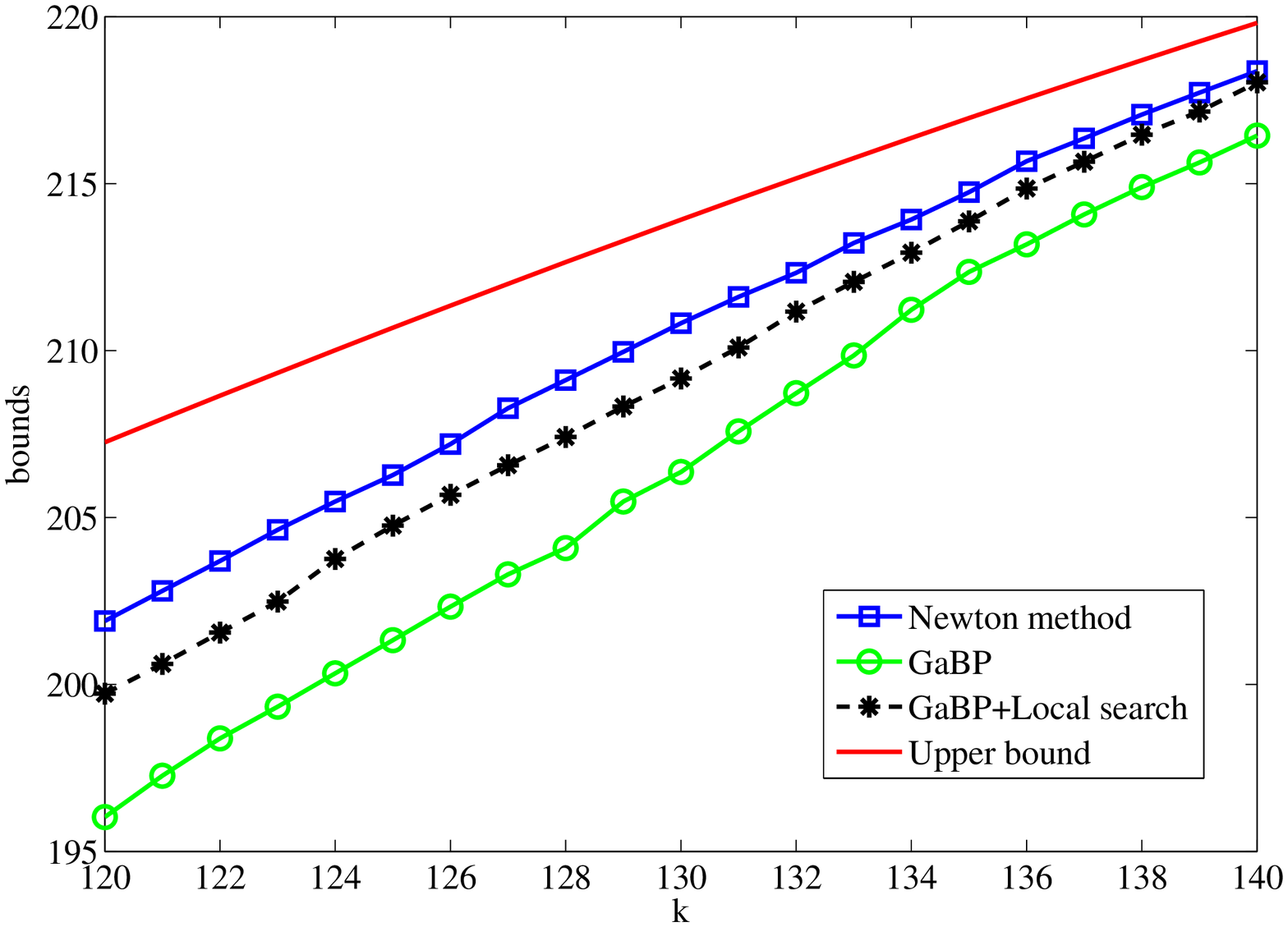}\\
\vspace{-3mm}
\end{tabular}
{\it Figure 1. Top: Synthetic data, bottom: Abilene data.}
\subsection{Abilene network data}
We have downloaded data with daily activity statistics of Abilene Internet2 backbone available 
from \cite{Abilene}. The data reported was collected between June 2006 to December 2006, 
total of $m=153$ days. In each day the average daily utilization of every Abilene network 
inbound and outbound links is given. Since there is a high correlation between inbound and 
outbound links, we have used only the inbound link activity. After filtering out links that where 
not active during the full examined period, we remained with $n=89$ links.   Our goal is to 
select the $k$ days that characterize normal network operation, for making a distinction with 
the other $m-k$ days that may indicate abnormal network behavior. Figure 1 (bottom) outlines the performance of the GaBP algorithm vs. the distributed Newton 
method. The $x$-axis presents $k$, the number of days selected. The $y$-axis represents 
the duality gap and the upper bounds respectively. As expected, the distributed Newton 
method has higher accuracy, but heaver computational overhead. As a sanity check, we have 
compared the number of error tickets opened each day and found a high correlation between 
them and the output of the distributed approximate algorithm. It is interesting to note that relative to the synthetic data set, the sensor selection problem becomes more difficult in the sense that the approximation algorithm is less accurate. This can be explained that the underlying distribution used by \cite{SS} is multivariate Gaussian with diagonal inverse covariance matrix, while real Internet traffic      typically does not come from a Gaussian distribution.

\section{Conclusion}
We have presented two distributed algorithms for solving the sensor selection problem. The 
first one is an accurate distributed version of the interior point method proposed in \cite{SS}. 
The second is a fast and light-weight approximation algorithm. Using simulations we have 
analyzed and compared to performance of both algorithms, using synthetic and real data 
collection from the Abilene Inetrnet2 backbone network. We believe there are several 
important applications to our novel construction for example in the area of quantitative database cleaning 
or distributed anomaly detection in communication networks.
\section*{Acknowledgment}
Danny Dolev is Incumbent of the Berthold Badler Chair in Computer Science. Danny Dolev was  supported in part by the Israeli Science Foundation (ISF) Grant number 0397373.
D. Bickson would like to thank Prof. Stephen Boyd from Stanford for interesting discussions.
Danny Bickson was partially supported by grants NSF IIS-0803333,
NSF NeTS-NBD CNS-0721591
and
DARPA IPTO FA8750-09-1-0141.\bibliographystyle{IEEEtran}   
\footnotesize
\bibliography{ISIT10-2}       

\end{document}